%% file: clustering.tex
\documentclass[iop,numberedappendix,appendixfloats]{emulateapj}
\usepackage{xspace}

\newcommand{\n}{\noindent}

\newcommand{\numOfLCs}{59}

\begin{document}

\title{Unsupervised clustering of Type II supernova light curves}
\keywords{}
\author{Adam Rubin and Avishay Gal-Yam}
\affil{Weizmann Institute of Science}

\begin{abstract}

As new facilities come online, the astronomical community will be provided with extremely large datasets of well-sampled light curves (LCs) of transient objects. This motivates systematic studies of the light curves of supernovae (SNe) of all types, including the early rising phase. We performed unsupervised k-means clustering on a sample of 59 R-band Type II SN light curves and find that our sample can be divided into three classes: slowly-rising (II-S), fast-rise/slow-decline (II-FS), and fast-rise/fast-decline (II-FF). We also identify three outliers based on the algorithm. We find that performing clustering on the first two components of a principal component analysis gives equivalent results to the analysis using the full LC morphologies. This may indicate that Type II LCs could possibly be reduced to two parameters. We present several important caveats to the technique, and find that the division into these classes is not fully robust and is sensitive to the uncertainty on the time of first light. Moreover these classes have some overlap, and are defined in the R-band only. It is currently unclear if they represent distinct physical classes, and more data is needed to study these issues. However, our analysis shows that the outliers are actually composed of slowly-evolving SN IIb, demonstrating the potential use of such methods. The slowly-evolving SNe IIb may arise from single massive progenitors. 

\end{abstract}

\maketitle

\section{Introduction}
There is a controversy in the literature regarding the division of SNe II into photometric sub-classes. The classical division into Type II-P and II-L events \citep{barbon_photometric_1979} based on the post-peak light curve has been supported by \cite{arcavi_caltech_2012} and \cite{faran_photometric_2014,faran_sample_2014}, but challenged by \cite{anderson_characterizing_2014} and \cite{sanders_toward_2015}. Until now, the focus has been on extracting characteristics from light curves (e.g. peak magnitude, post-peak decline rate, plateau length), and searching for separation into populations. However, it is also known that some SNe have strikingly similar light curves, and the determination of which characteristics to compare is not straightforward. 

New surveys such as the Zwicky Transient Facility \citep[ZTF;][]{bellm_zwicky_2014} and the Large Synoptic Survey \citep[LSST;][]{ivezic_lsst:_2008} will provide the astronomical community with large datasets of well-sampled light curves (LCs). The rolling nature of these surveys will provide good coverage of the poorly studied early phases of SNe. However, there is and will remain a permanent lack of resources for spectroscopic classification of the transients found. Therefore it is of utmost importance to develop tools for the study and classification of SN LCs based on their photometry alone.

Here we attempt to divide the sample of 57 LCs presented in \cite{rubin_type_2015}---in addition to two SN LCs which were originally included in that sample but subsequently removed---into classes of similar LC shape by using the unsupervised clustering algorithm K-Means on the LCs directly (as opposed to extracted parameters). This algorithm is one of the simplest both conceptually and algorithmically. Unsupervised clustering on SN characteristics (using a different technique) has been previously used for Type Ia SNe by \cite{benetti_diversity_2005}. K-means has been used for many applications in the past, however its use on SN light curves has not yet been attempted---perhaps due to the sparse sampling of typical SN LCs.

The \cite{rubin_type_2015} sample has the advantage of containing LCs with very well constrained times of first light, allowing us to utilize the early-time LC behavior which has previously been unaccessible. Here we explore how the sample divides into classes, and explain this technique's limitations and pitfalls when it is applied to SN LCs.

\section{Analysis}

We used the light curve sample presented in \cite{rubin_type_2015}, and added two SNe (iPTF13blq and iPTF14bas; Arcavi et al. 2016 \emph{in prep.}) which were originally in that sample, but were removed for reasons explained below and discussed in Section \ref{sec:IIb}. We smoothed the LCs using the algorithm described in \cite{rubin_type_2015}. The \cite{rubin_type_2015} algorithm did not fit iPTF14bas well around peak, and we found that a smoothing spline performed much better. The smoothed LCs of both iPTF13blq and iPTF14bas are presented in appendix Figure \ref{fig:13blq_14bas_smoothed}.

We used the K-means++ \citep{arthur_k-means++:_2007} algorithm, implemented as \emph{kmeans} in Matlab, to automatically divide our light curves into classes. We interpolated our smoothed light curves to measure the magnitude between rest frame days 1 and 30, with uniform spacing of 0.1 days. This is equivalent to treating the data set as \numOfLCs{} observations (the number of SNe in the sample) of 291 highly correlated parameters (relative R-band magnitude at 291 times). Data from $t<1$ days was ignored because it contained information that was interpolated between the last limit and the first detection. 

K-means is an iterative algorithm that searches for proximity in multi-dimensional spaces. It is necessary to specify the number of clusters beforehand. The algorithm operates as follows:
\begin{enumerate}
\item Randomly select a starting point for each cluster in the multi-dimensional space.
\item Classify all points according to their closest cluster. \label{item:classify}
\item Calculate a new centroid for each cluster based on the classified points.\label{item:newCentroid}
\item Repeat \ref{item:classify} and \ref{item:newCentroid} until convergence.
\end{enumerate}   

\n We used the simplest distance metric---sum of squares---to determine proximity. On the one hand this is reasonable because all of our observations are in magnitude differences. On the other hand, the rise is much shorter in duration than the decline/plateau. Therefore late time data can dominate the division. Limiting our analysis until day 30 after first light gives similar weight to both the rise and the decline of the LCs. There are various techniques for determining the number of clusters, however in real-world cases the correct number of clusters is neither trivial not objective. We found that dividing our sample into four clusters gave an interesting subdivision of the data. K-means was initialized randomly 100 times to eliminate sensitivity to the arbitrary starting point of the algorithm.

In order to visualize our results, we also performed principal component analysis (PCA). PCA finds directions in multi-dimensional sample space (in descending order) along which the variance is maximal. In other words, it finds eigenvector LCs which give the highest variance when the sample LCs are projected onto them. Defining our $m\times n$ matrix of LCs (where m is the number of SNe and n is the number of parameters) as $\bf{K}$, and our PCA eigenvectors as $\vec{p}_i$, the principal components $\vec{c}_i$ are given by

\begin{equation}
	\vec{c}_i = \textbf{K} \cdot \vec{p}_i
\end{equation}

\n Note that $\vec{c}_i$ is a vector of the $i^{th}$ components of all of the LCs in the sample.

\section{Results}
\begin{figure}[ht]
\centering
\includegraphics[width=1\columnwidth]{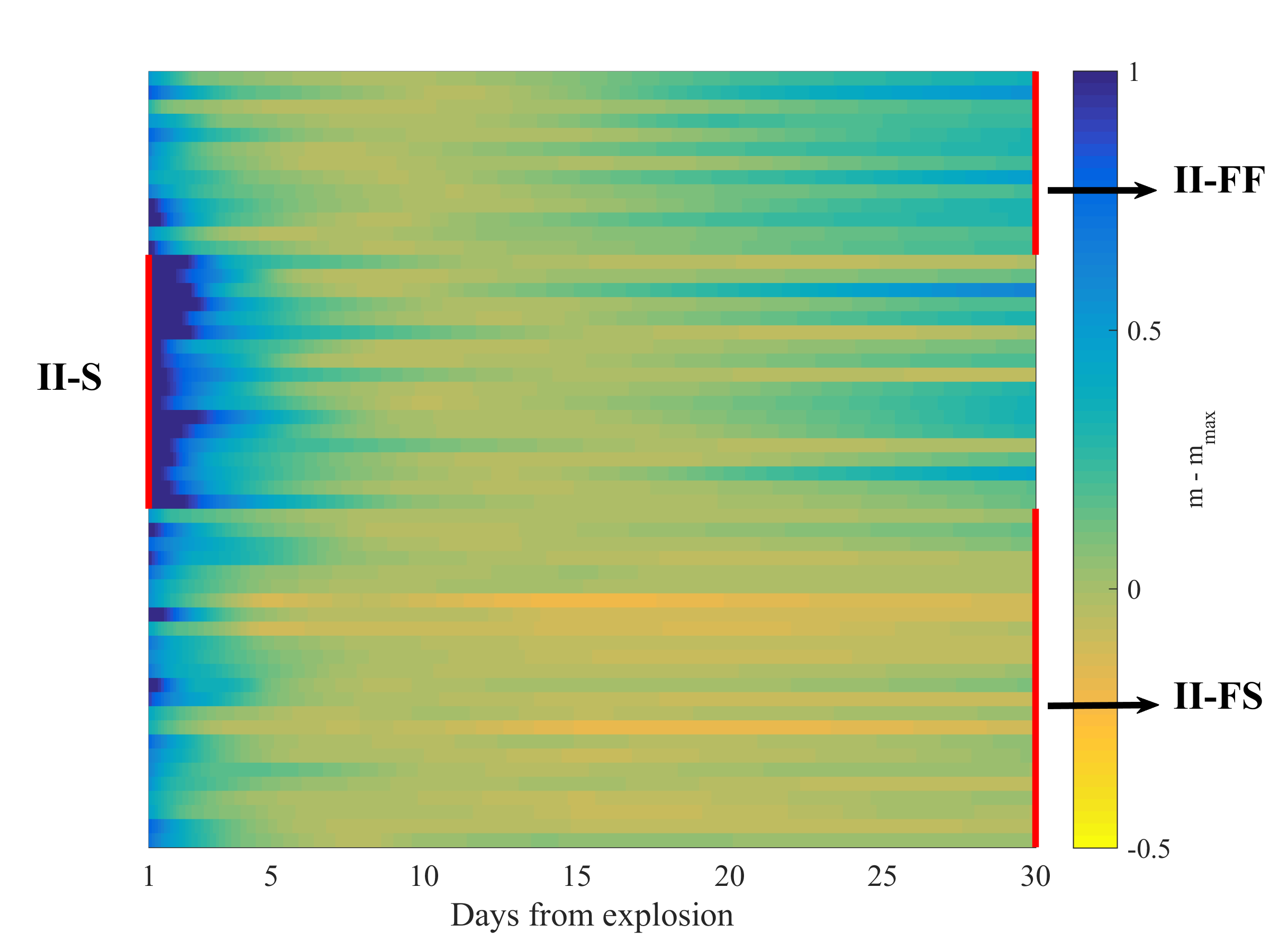}
\caption{Visual plot of the clusters. Color represents the magnitude (normalized to the peak). Each row is a SN, where the SNe have been ordered by cluster. The red line highlights the cluster locations. We can clearly see a separation into three clusters.}
\label{fig:ClusterColorDiagram}
\end{figure}

\begin{figure*}[ht]
\centering
\includegraphics[width=0.85\textwidth]{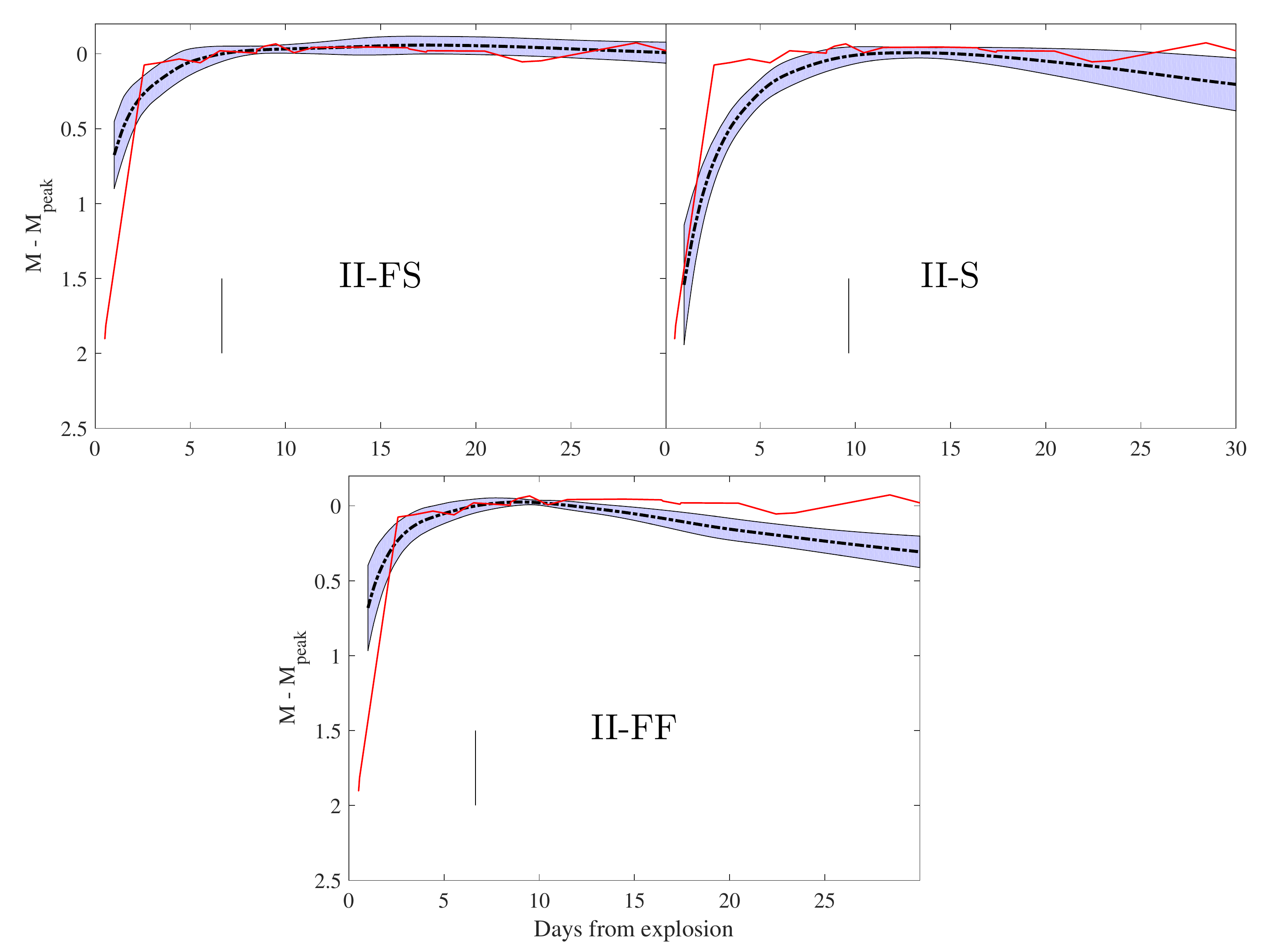}
\caption{K-means clusters found using R-band light curves sampled from day 1 to 30. The median rise-time of each cluster has been marked with a vertical line. Type II-P SN 2005cs \citep{pastorello_sn_2009} is superimposed, and  fits only the II-FS category. This is the only well-studied SN II-P with good sampling of the rise.}
\label{fig:Clusters}
\end{figure*}

\begin{figure}[ht]
\centering
\includegraphics[width=1\columnwidth]{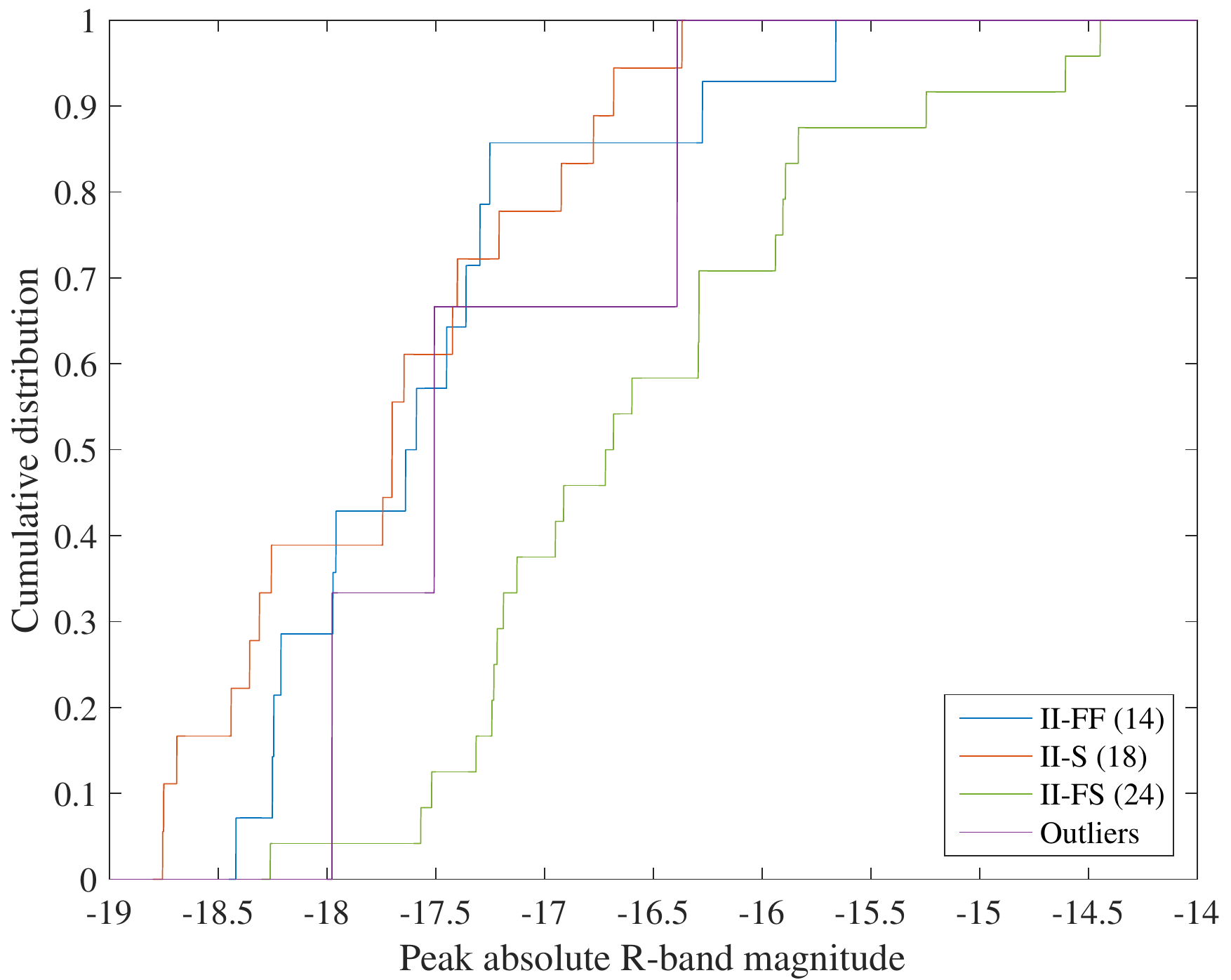}
\caption{Cumulative distribution function of the peak absolute R-band magnitude for each of the clusters. The number of objects in each cluster are given in parenthesis. The II-FS cluster is enriched in lower-luminosity events, and does not come from the same population as the II-FF and II-S clusters at the 95\% significance level (based on a KS test).}
\label{fig:ClusterHistograms}
\end{figure}

\begin{figure}[ht]
\centering
\includegraphics[width=1\columnwidth]{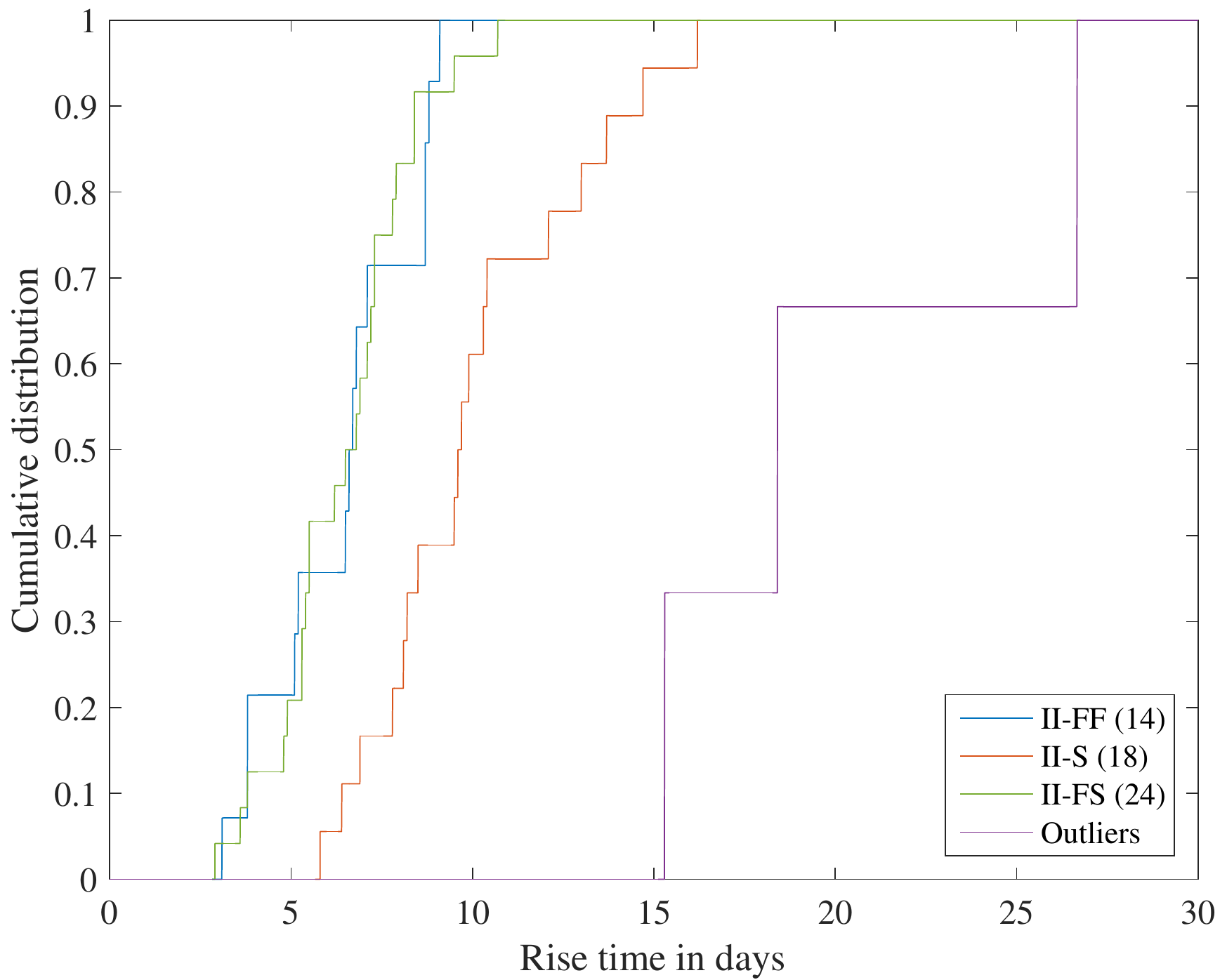}
\caption{Cumulative distribution function of the rise-time for each of the clusters. The number of objects in each cluster is given in parenthesis. The II-S cluster is not consistent with having the same parent population as the II-FF and II-FS clusters at the 95\% significance level (based on a KS test).}
\label{fig:riseTimeHistogram}
\end{figure}

\begin{figure}[ht]
\centering
\includegraphics[width=1\columnwidth]{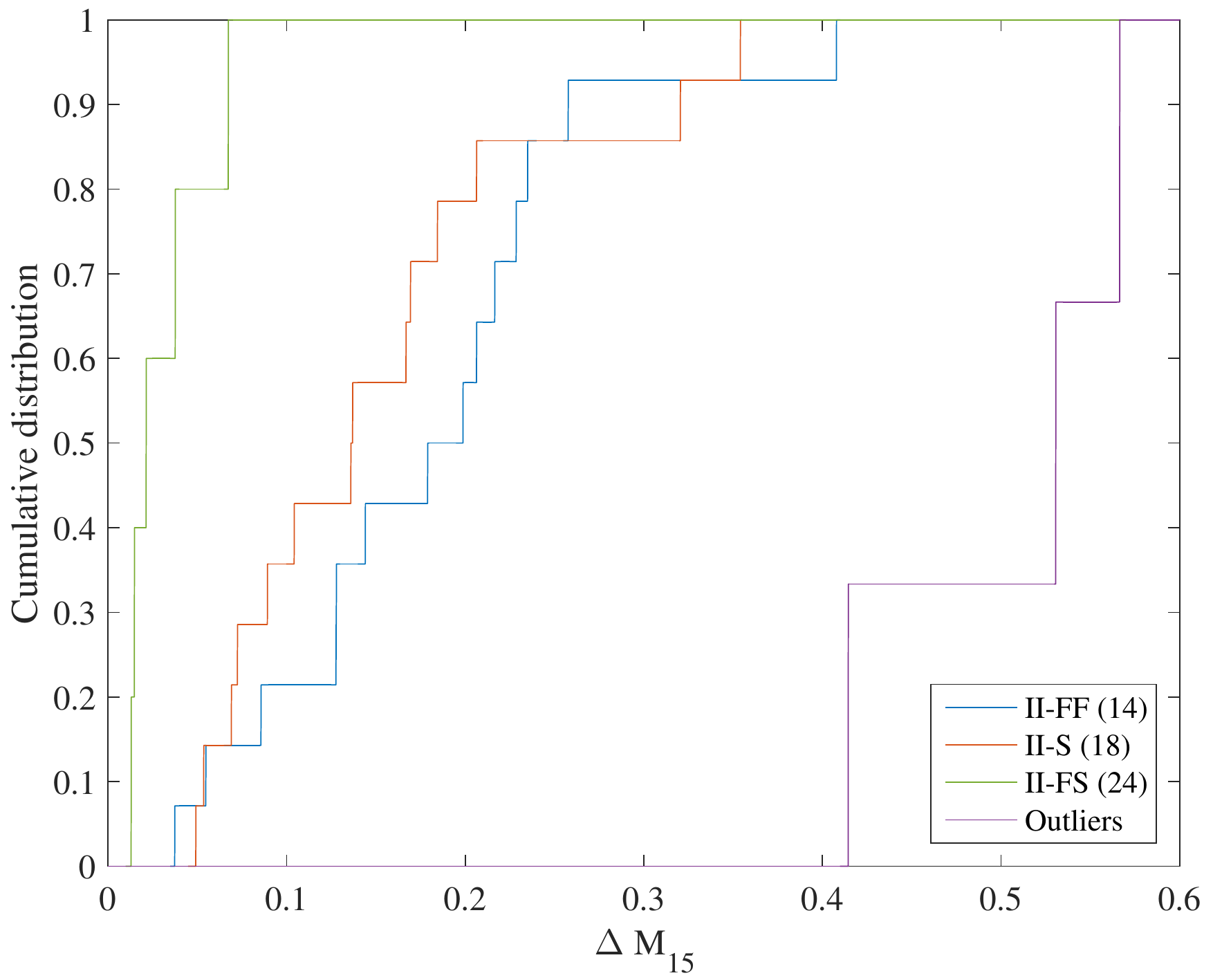}
\caption{Cumulative distribution function of $\Delta m_{15}$ for each of the clusters. The number of objects in each cluster is given in parenthesis. The II-FS cluster is not consistent with having the same parent population as the II-FF and II-S clusters at the 95\% significance level (based on a KS test).}
\label{fig:deltaM15Histogram}
\end{figure}

We found that dividing our SN light curve sample into four clusters provides an intriguing physical interpretation. One cluster contained only two events (iPTF13blq and iPTF14bas) which we identify as outliers. After the analysis was completed using four clusters we also identified a third outlier (iPTF14ajq), which was subsequently also identified by the algorithm once the number of clusters was raised to five. We removed these three events and discuss them separately in Section \ref{sec:IIb}. We found that the remaining LCs were best divided into three clusters. A color visualization of the light curves sorted by cluster is presented in Figure \ref{fig:ClusterColorDiagram}. The rise time, which is uniquely available for this sample, plays a major role. Two clusters which have fast rise times (II-F) are separated by their decline phase. One cluster has a fast rise and slow decline (II-FS; similar to a II-P), while the other has a fast rise but fast decline (II-FF; similar to a II-L). The remaining cluster is separated by its slow rise (II-S). The II-FS, II-FF, and II-S clusters are of similar size. The centroid LCs and the standard deviations of each cluster are shown in Figure  \ref{fig:Clusters} and given in the online material (the structure of the table is given in Table \ref{tab:meanStdTablePreview}). Histograms showing the cumulative distribution of peak magnitude, rise-time, and $\Delta m_{15}$\footnote{Defined as the difference in magnitude between the peak and 15 days after the peak.} within each cluster are shown in Figures \ref{fig:ClusterHistograms}, \ref{fig:riseTimeHistogram}, and \ref{fig:deltaM15Histogram} respectively. Note that the II-FS cluster does not come from the same population as II-FF and II-S in both peak magnitude and decline rate at the 95\% significance level according to a Kolmogorov-Smirnov (KS) test. The II-S cluster does not come from the same population as the II-FF and II-FS clusters in rise-time at the 95\% significance level according to a KS test.

\begin{figure}[ht]
\centering
\includegraphics[width=1\columnwidth]{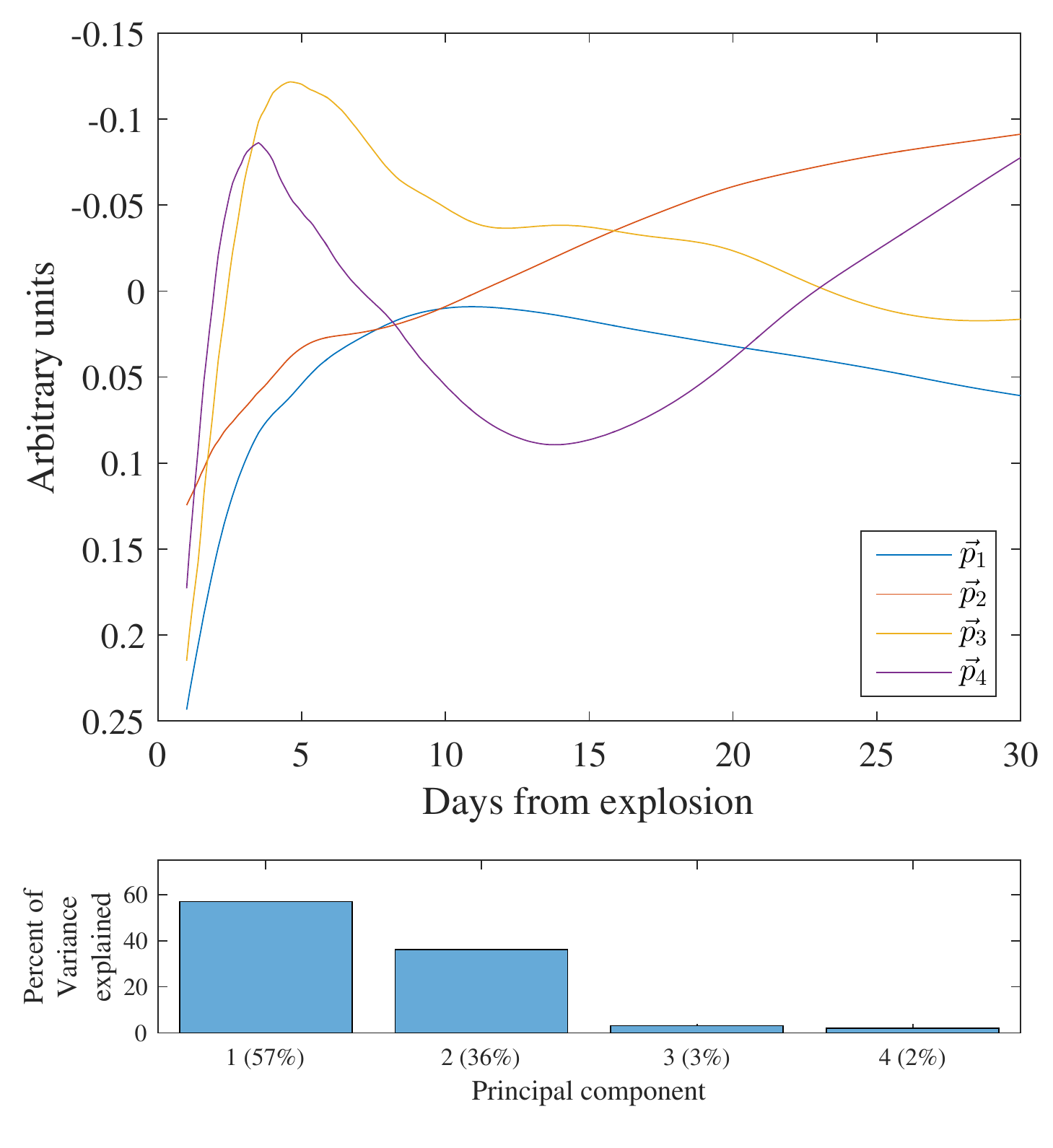}
\caption{Top panel: the first four PCA eigenvectors. Bottom panel: percent of the variance explained by each component. The first two components explain 90\% of the observed variance.}
\label{fig:pcaEigenvectors}
\end{figure}

\begin{figure}[ht]
\centering
\includegraphics[width=1\columnwidth]{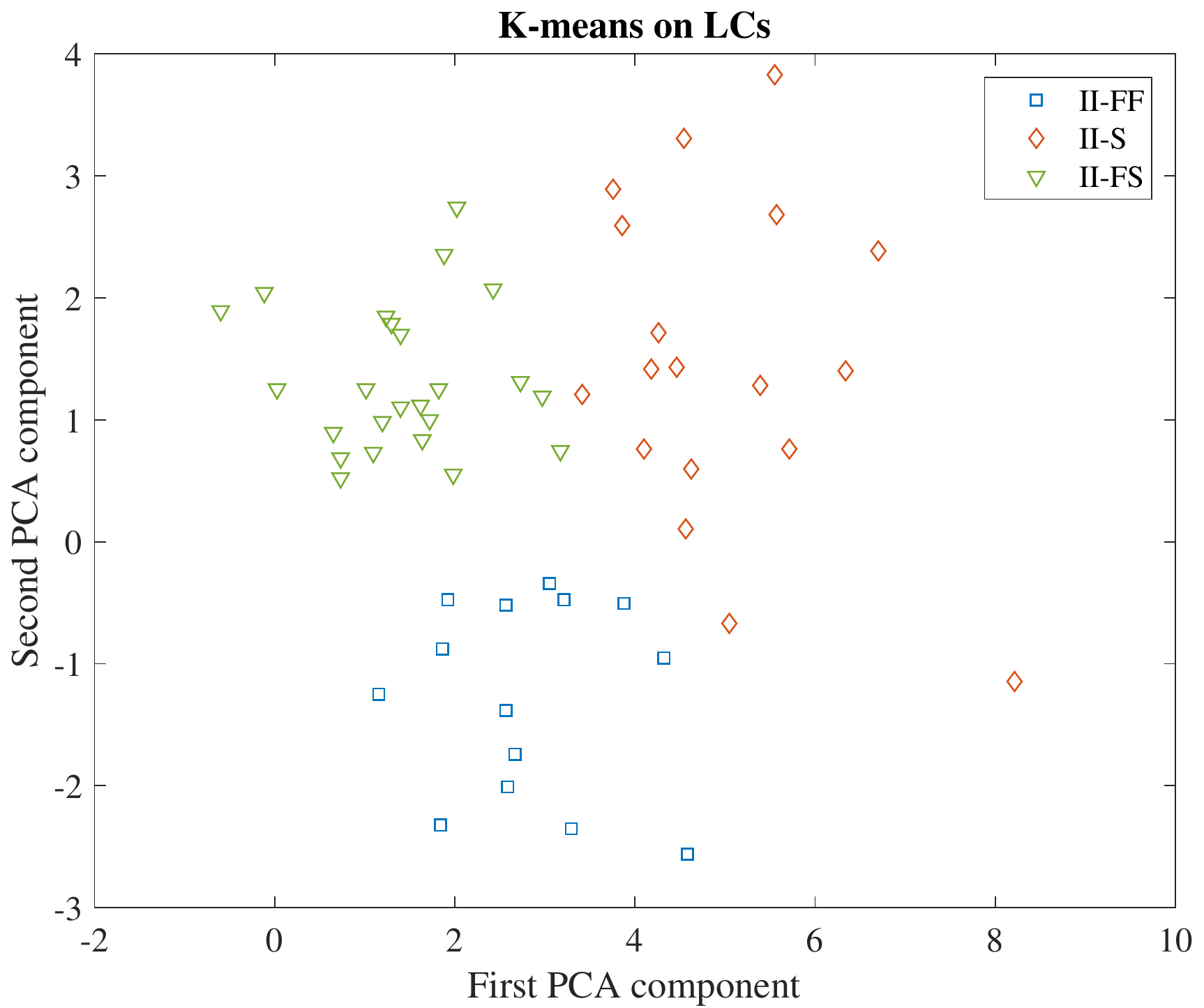}
\includegraphics[width=1\columnwidth]{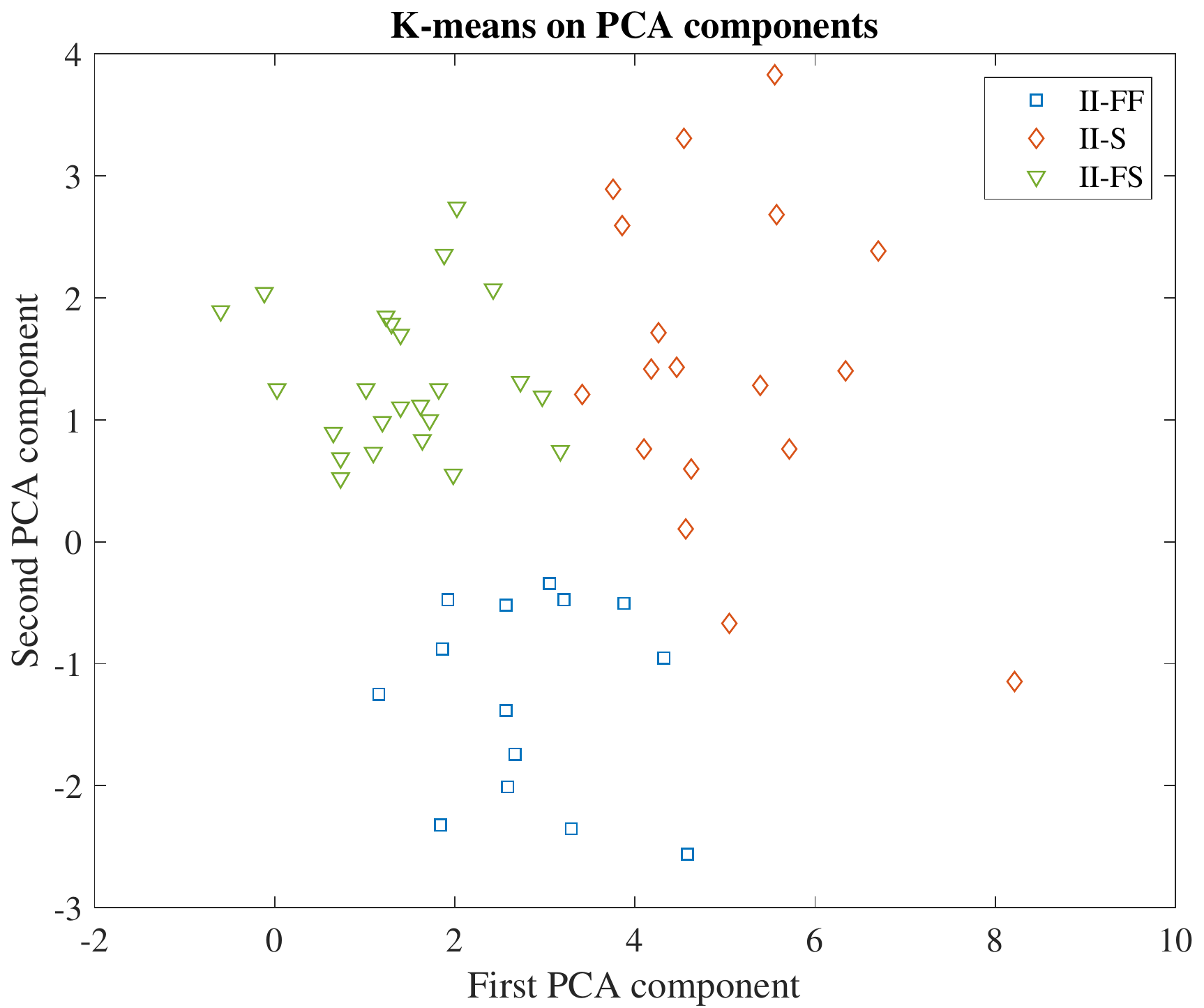}
\caption{First two PCA components colored by their cluster association. Top: objects associated by applying K-means directly on the LCs. Bottom: objects associated by applying K-means on the first two PCA components. The full interpolated LCs (until day 30) encode similar information as the first two PCA components.}
\label{fig:compToPCA}
\end{figure}

\begin{figure}[ht]
\centering
\includegraphics[width=1\columnwidth]{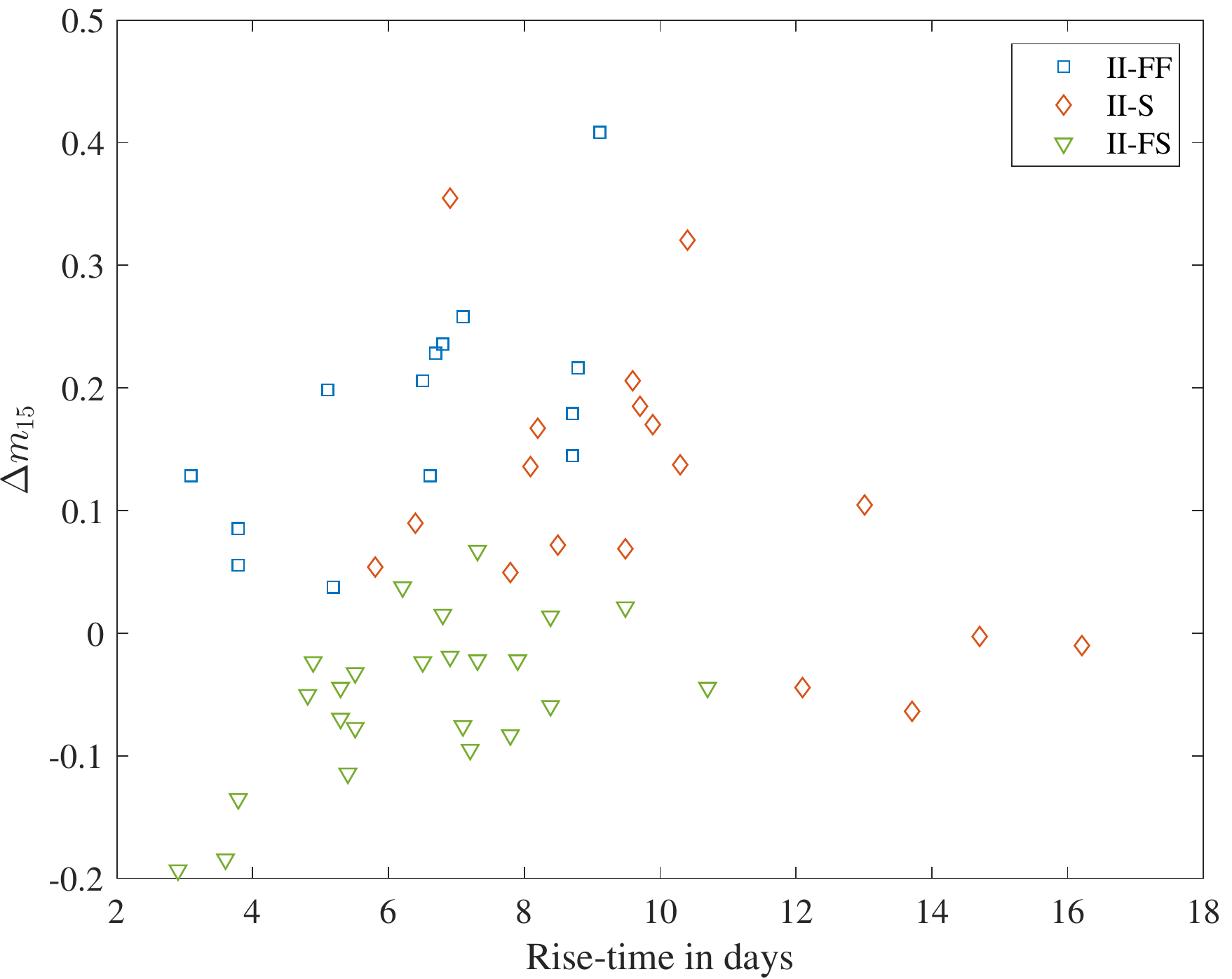}
\caption{$\Delta m_{15}$ as a function of rise-time colored by cluster association.}
\label{fig:triseVdeltaM15ColoredByCluster}
\end{figure}

The first four PCA eigenvectors are shown in Figure \ref{fig:pcaEigenvectors}. The first two components describe most ($90\%$) of the variance between light curves. They are a slow rising, and fast rising fast declining light curve. Figure \ref{fig:compToPCA} shows that applying K-means clustering on the interpolated light curves is equivalent to performing clustering on the projection of the LCs on the first two PCA eigenvectors. Moreover, we show that applying K-means directly on the PCA projections divides our sample into identical clusters. Figure \ref{fig:triseVdeltaM15ColoredByCluster} shows that the algorithm is not clustering based on rise-time and decline rate. Clustering based on the LC morphology is not equivalent to clustering based on rise-time and decline rate.

\begin{figure}[ht]
\centering
\includegraphics[width=1\columnwidth]{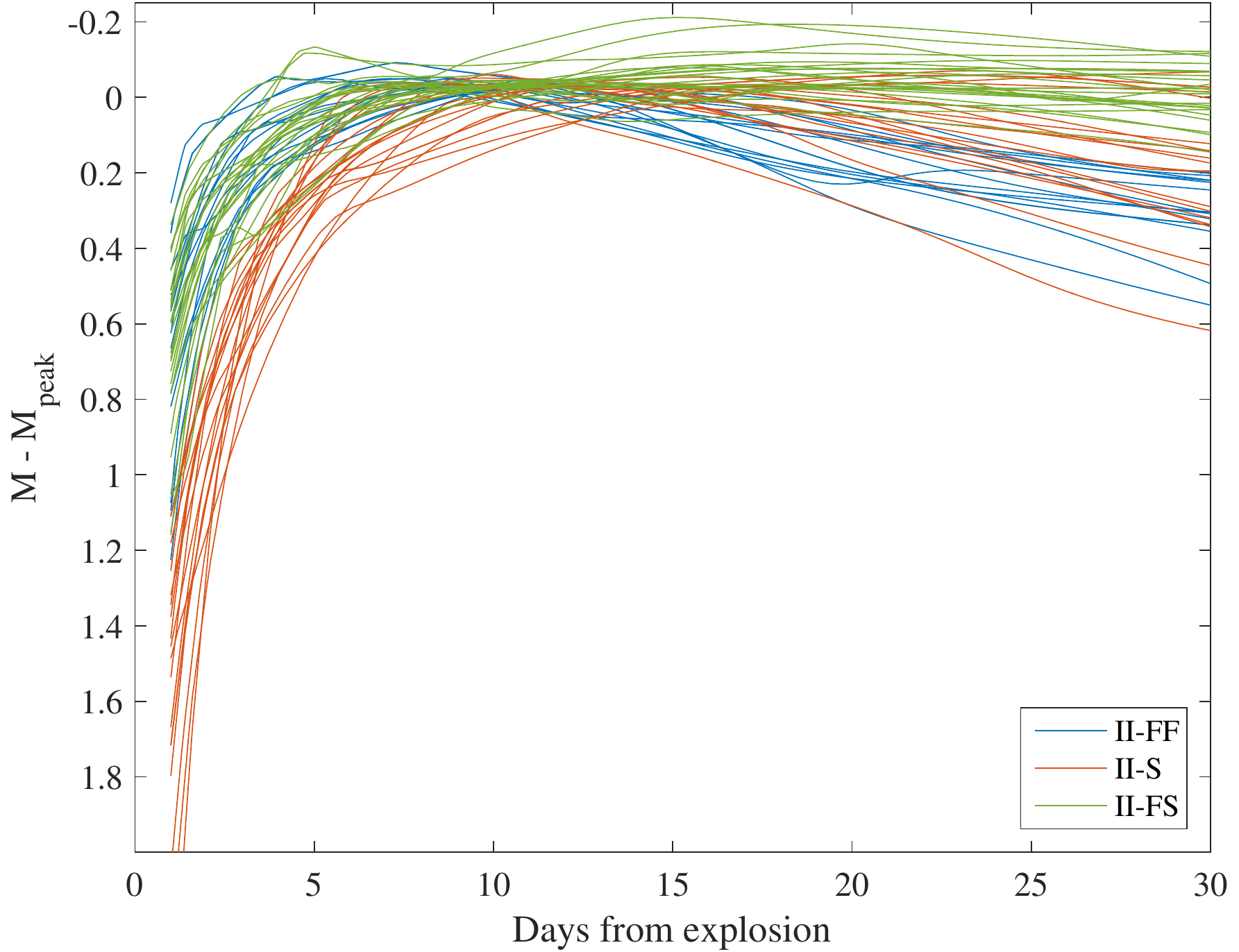}
\caption{The sample LCs colored by cluster.} 
\label{fig:clusterLCs}
\end{figure}

\begin{figure}
\centering
\includegraphics[width=1\columnwidth]{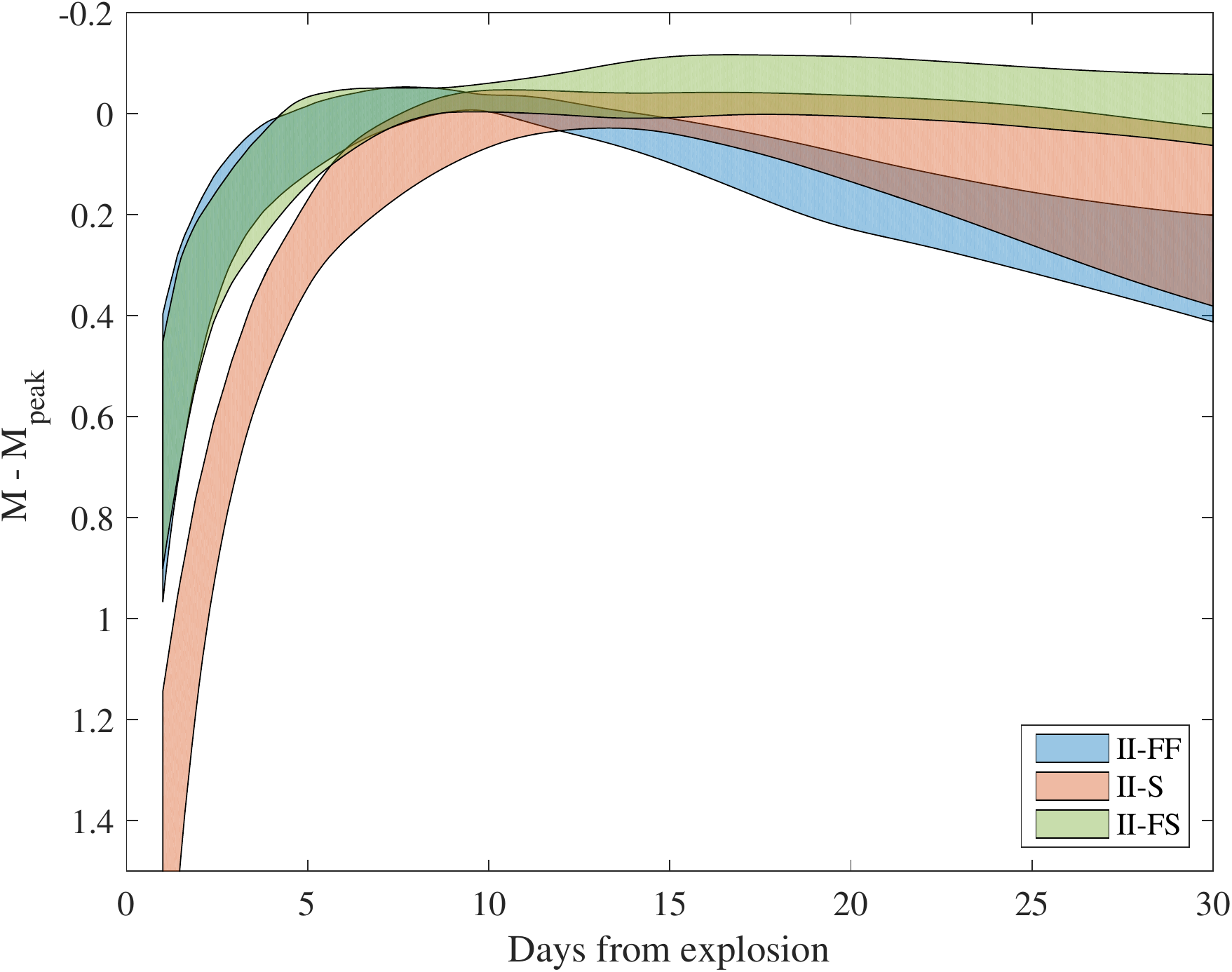}
\caption{Plots of the cluster $1-\sigma$ limits. Notice that II-FF and II-FS are disjoint at late times, and the two are weakly disjoint from the II-S at early times. The slow-risers bridge the gap at late time between II-FF and II-FS.}
\label{fig:clusterComparison}
\end{figure}

\begin{deluxetable}{ccccccc}
        \tablewidth{0pt} 
        \tablecolumns{7}
        \tablecaption{Mean and standard deviation of each cluster. The full table is given in the online material. \label{tab:meanStdTablePreview}}
        \tablehead{  
                \colhead{}& \multicolumn{2}{c}{II-FS}&\multicolumn{2}{c}{II-FF} &\multicolumn{2}{c}{II-S}  \\
                \colhead{Phase} &
                \colhead{Mean} & \colhead{$\sigma$} &
		\colhead{Mean} & \colhead{$\sigma$} &
		\colhead{Mean} & \colhead{$\sigma$} \\
		\colhead{Days} & \multicolumn{6}{c}{$M - M_{peak}$}
        }
        \startdata
        \input{meanStdTablePreview.txt}
	\enddata
\end{deluxetable}

\section{Caveats}

We draw the reader's attention to the following caveats and issues we observed during this analysis.

\begin{figure*}[ht]
\centering
\includegraphics[width=1\textwidth]{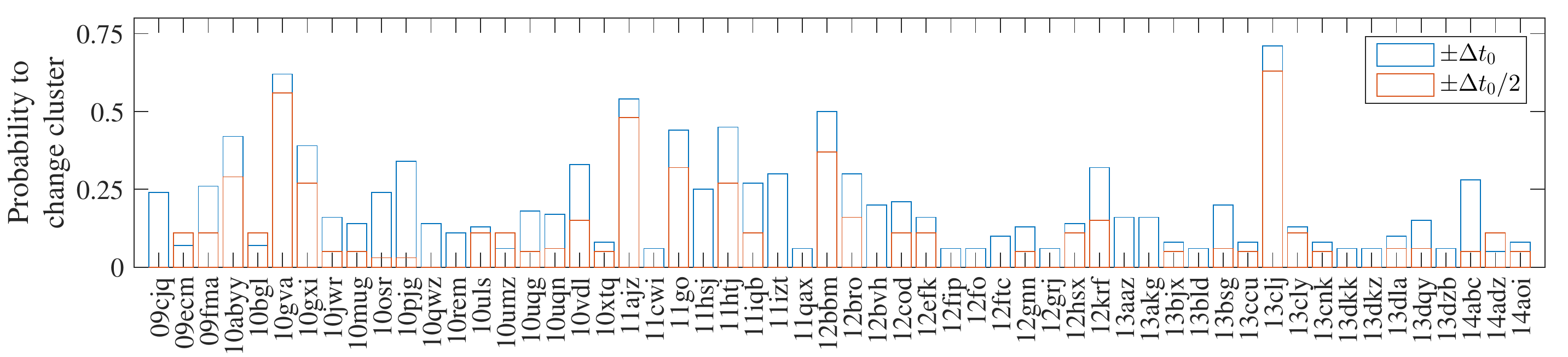}
\caption{Probability of each object to change cluster when LCs were shifted randomly in time between $[-\Delta t_0, +\Delta t_0]$ and $[-\Delta t_0/2, +\Delta t_0/2]$.}
	\label{fig:probToChangeCluster}
\end{figure*}

\begin{figure}[ht]
\centering
\includegraphics[width=1\columnwidth]{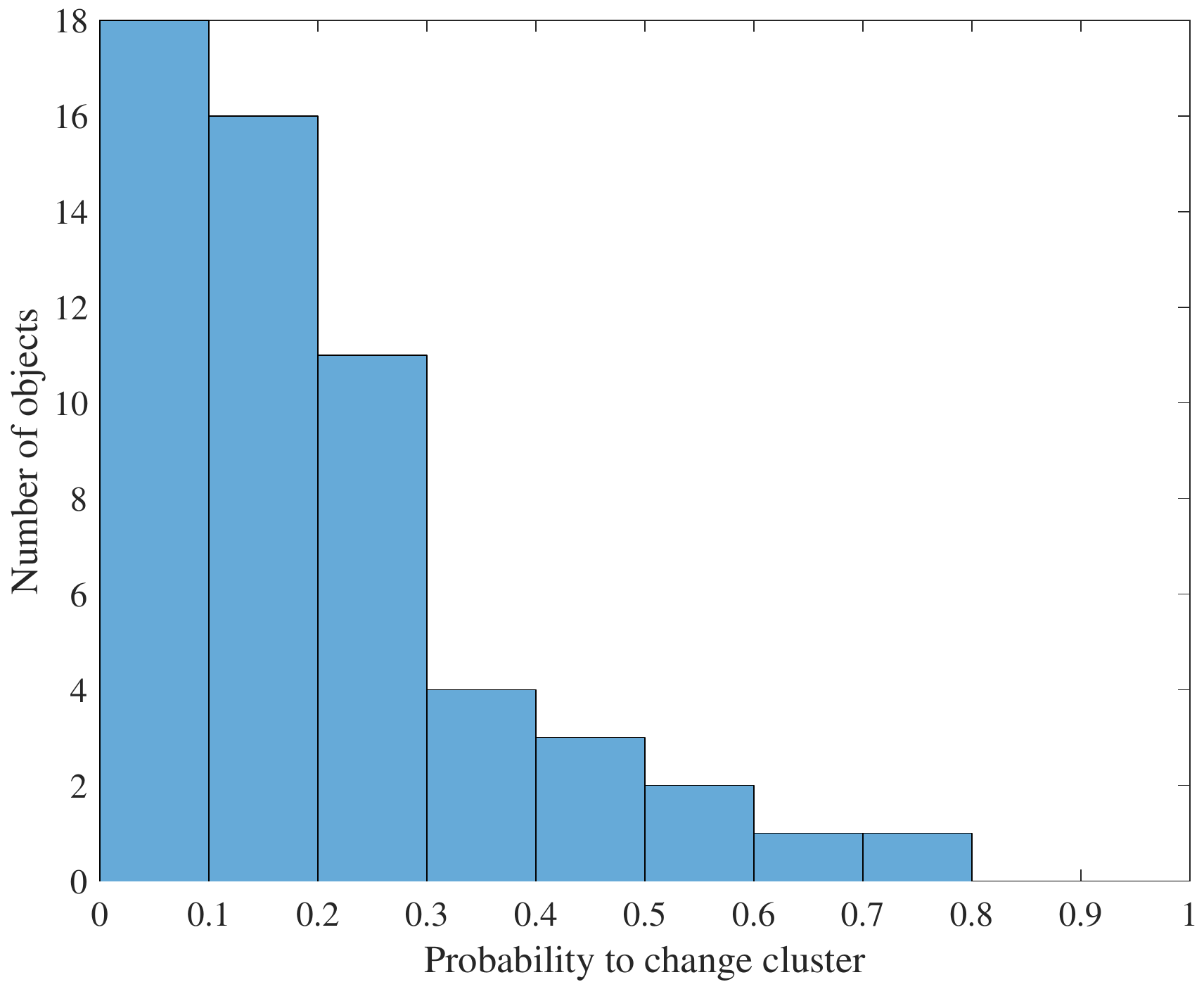}
\caption{Histogram of the probability to change cluster when LCs were shifted randomly in time between $[-\Delta t_0, +\Delta t_0]$.}
	\label{fig:histProbToChangeCluster}
\end{figure}

\begin{figure}[ht]
\centering
\includegraphics[width=1\columnwidth]{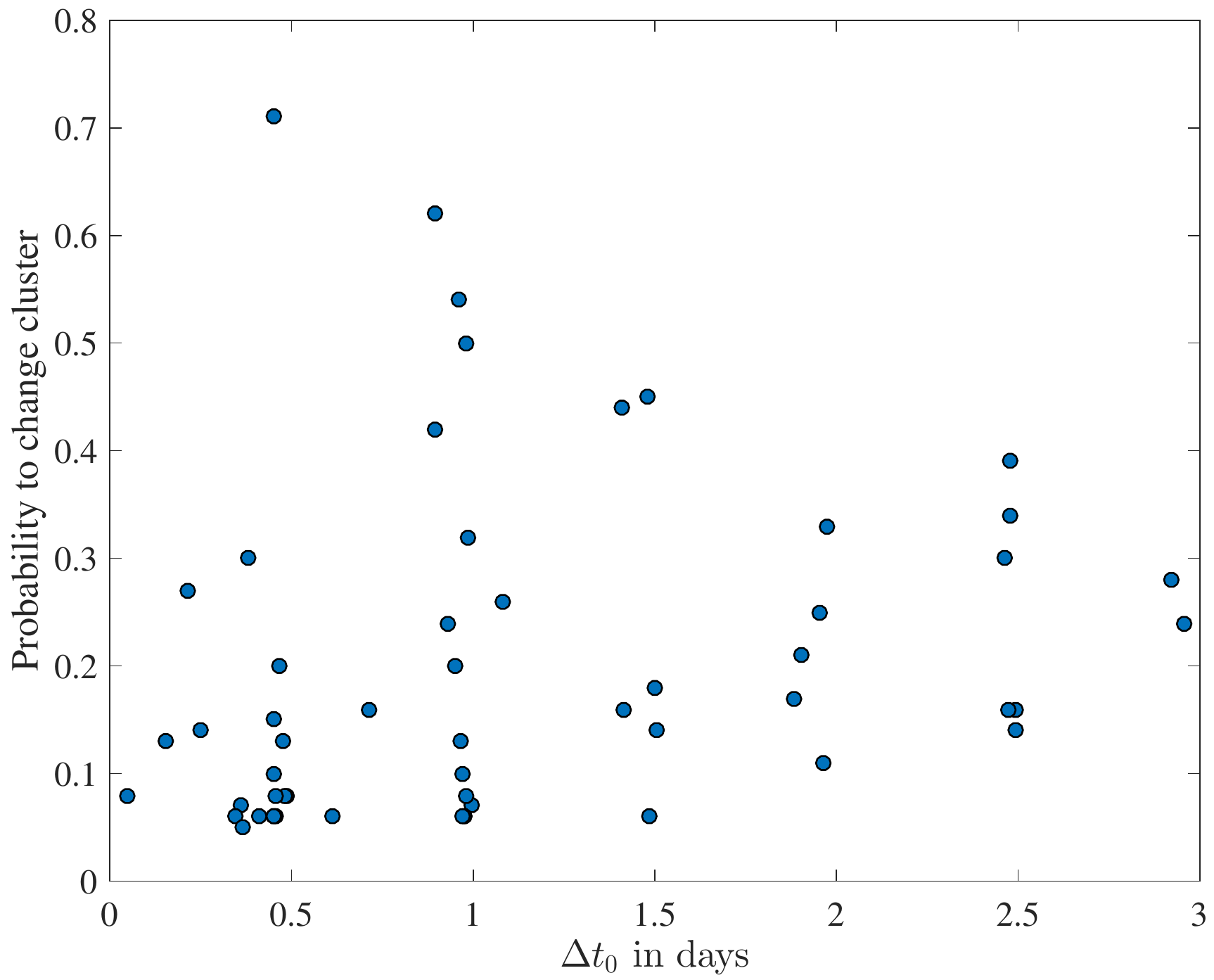}
\caption{Probability to change cluster plotted against the uncertainty on the time of first light. No correlation is apparent.}
\label{fig:probToChangeClusterVst0Error}
\end{figure}

\begin{itemize}
	\item The clustering is sensitive to the determination of the time of explosion. Running 100 simulations where we shifted each light curve randomly in time over the range $[-\Delta t_0, +\Delta t_0]$ we found that 34 objects had a probability of changing cluster less than 20\%. This is extremely conservative, because for most events there are measurements during the rise, making a random uniform distribution an extreme overestimation of the uncertainty. Figure \ref{fig:probToChangeCluster} shows the probability of each SN to change cluster and Figure \ref{fig:histProbToChangeCluster} shows how these probabilities are distributed. Some objects are more sensitive to these shifts than others. Despite what one might expect, the probability to change cluster does not appear to be correlated with the uncertainty on the time of explosion (see Figure \ref{fig:probToChangeClusterVst0Error}).
	\item An optimal way of weighting the full light curve (including information beyond day 30) is still lacking.
	\item We do not have a robust objective metric to determine the number of clusters. We did not find the \cite{calinski_dendrite_1974} test (which maximizes the inter-variance while minimizing the inner-variance) to be robust. In some cases the optimal number of clusters according to the test was 56. An improved metric is clearly necessary.
	\item We normalize the LCs by their peak luminosity. This may be reasonable given our uncertainties regarding host galaxy extinction, and our objective of finding SNe that have very similar LC shapes, but very different peak luminosities. However, much of the physics is encoded in the peak luminosity, and with a larger sample it may be more meaningful to explore clustering while taking peak luminosity into account. 
	\item Our clustering is performed in R-band. It is well known that LC shapes can vary in different filters, and R-band may not be the optimal filter in which to perform such an analysis.
\end{itemize}

\section{Discussion}

Several recent works explore the morphology of large samples of Type II light curves. \cite{arcavi_caltech_2012,anderson_characterizing_2014,faran_photometric_2014,faran_sample_2014} and \cite{sanders_toward_2015} tried to assess whether II-P and II-L light curves actually form two disjoint classes, or are rather a continuous class. They all examined post-maximum morphology. \cite{arcavi_caltech_2012} and \cite{faran_photometric_2014, faran_sample_2014} found that II-P and II-L are disjoint in R-band, but may have suffered from small number statistics. Studies of larger samples \citep{anderson_characterizing_2014,sanders_toward_2015} led their authors to the conclusion that the decline rates, peak magnitudes, and plateau lengths are all continuous properties, with no clear separation into disjoint populations. \cite{anderson_characterizing_2014} was particularly careful in fitting the post maximum decline rate separately from the plateau decline rate, and found these quantities to be continuous in a sample of 116 V-band light curves. More recently \cite{rubin_type_2015} presented a sample of 57 very well sampled LCs and did not observe any clear discontinuities in the LC properties including the rise-time.

We have, for the first time, performed unsupervised clustering on Type II light curves. Importantly, our method takes into account the full morphology until day 30, including rise time data. Our division into four SN light curve classes leads us to several conclusions. First, while the rapidly rising SNe appear to naturally separate into two populations, divided by their slow/fast declines (II-FS and II-FF, which resemble II-P and II-L) the late-time morphology of the slow rising events (II-S, Figure \ref{fig:Clusters}) does not appear to be very different from the classic II-P/II-L classes. Therefore the previous classes of II-P and II-L may have had a large number of interlopers from this slow rising population---perhaps this is the source of the continuum behavior observed in the previous works mentioned. While there is some excess at longer rise-times, our sample is consistent with originating from a single population.

Figure \ref{fig:clusterLCs} overplots these three clusters, while Figure \ref{fig:clusterComparison} overplots the three centroid clusters and their standard deviations. It is clear from the figure that the fast-rise/slow-decline (II-FS, Figure \ref{fig:Clusters}) and fast-rise/fast-decline (II-FF, Figure \ref{fig:Clusters}) are disjoint at late times, but the II-S class bridges the gap---possibly inducing the observed continuum in previous works. 

Examining Figure \ref{fig:ClusterHistograms} we see that the II-FS cluster contains lower luminosity events, while the remaining clusters are comprised of brighter SNe on average (recall that the clustering does not take into account peak magnitude). This excess of faint objects is the slow/faint SN II-P population (e.g. SN 2005cs and PTF10vdl). Using a KS test we find that the II-FS cluster does not come from the same population as the II-FF and II-S clusters in peak magnitude at the 95\% significance level. The fact that the II-FS cluster differs in an observable other than LC shape may indicate that there is a physical meaning for the clusters found. 

Our comparison to PCA is instructive. We found that using the full LC information is equivalent to using just two parameters, the projections of the LCs on the first two principal components. This may have far reaching ramifications, and may indicate that Type II LCs of this quality are a two parameter problem. Including late time data, and the peak magnitude will add some complexity.

It is important to note that the classes identified in this work have some overlap, are limited to R-band, and it is unclear if they represent distinct physical classes. We have found some sensitivity to the determination of the time of explosion. Moreover, the sensitivity does not appear to be correlated with the uncertainty on the time of explosion, making it unclear if higher cadence photometry would improve the results. 

These clusters may be useful in photometric identification of SNe, or the completion of gaps in photometric coverage. Our discovery of a spectroscopically distinct class in an unsupervised fashion encourages further work. However the true test of these results will be the application of the this analysis on an independent sample of similar size and quality. As photometric surveys begin to supply the community with large data sets of similarly sampled LCs, it will likely become necessary to employ unsupervised techniques to identify and classify transients. 

\section{Outliers: Type II\MakeLowercase{b} SN\MakeLowercase{e} from massive progenitors}
\label{sec:IIb}

Examining the outliers more closely, we found that two of them (iPTF13blq and iPTF14bas) have spectra that are consistent with a IIb classification, although their light curves are unusually slowly-rising. This was the reason they were removed from the \cite{rubin_type_2015} sample after performing the analysis presented in this work. We present their spectra with similar IIb spectra superimposed in Appendix Figure \ref{fig:II-E_Classification}. Only later did we realize that iPTF14ajq is also an outlier, and we confirmed that the algorithm identifies this by increasing the number of clusters to five, which clustered iPTF13blq and iPTF14ajq together, leaving iPTF14bas as a cluster alone. Returning to \cite{rubin_type_2015} we find that the spectrum of iPTF14ajq suffered heavy galaxy contamination. While it is conclusively a Type II SN, we cannot rule out a IIb classification based on the only available spectrum. All three spectra are available on WISeREP \citep{yaron_wiserep_2012}.

Type IIb (as well as Type Ib and Ib/c) originate from stripped envelope progenitors. While the stripping mechanism is the subject of some debate, the two main possibilities are stripping via binary interaction \citep{podsiadlowski_presupernova_1992}, and stripping due to episodic mass loss or to stellar winds. While the low ejecta masses and high rates of SNe IIb \citep[19-25\% by volume;][]{li_nearby_2011} may point towards the binary interaction channel, it is expected that at least some SNe IIb will result from single massive stars that are stripped by winds. These are expected to produce a large ejecta mass, leading to a long-rising LC. SN 2009jf is an example a related event, where a massive star was almost completely stripped of its hydrogen envelope, leading to a Ib classification. \cite{valenti_sn_2011} and \cite{sahu_optical_2011} found that SN 2009jf was most likely a Type Ib SN originating from a massive single-star progenitor with $\rm{M_{ZAMS}} \gtrsim 20-30$ M$_\odot$. 

\begin{figure}[ht]
\centering
\includegraphics[width=1\columnwidth]{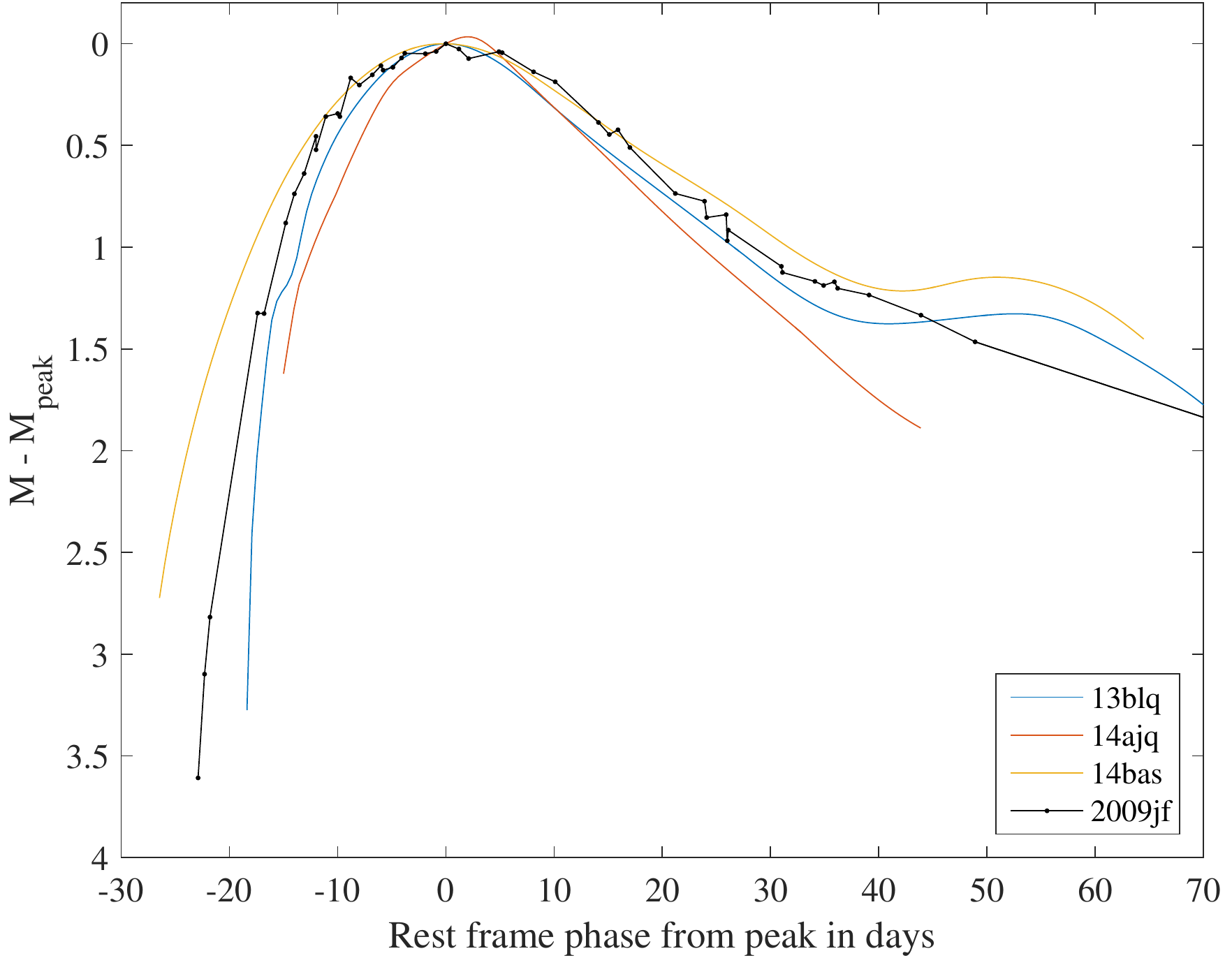}
\caption{Comparison of iPTF13blq, iPTF14bas, and iPTF14ajq to SN 2009jf.}
\label{fig:II-E_compare_2009jf}
\end{figure}

In Figure \ref{fig:II-E_compare_2009jf} we compare the outliers to SN 2009jf and find that the light curves are strikingly similar. We speculate that iPTF14ajq is in fact a IIb and suggest that these extremely slowly-evolving IIb's originate from massive stripped envelope stars. They are also relatively luminous, two with absolute peak magnitudes less than -17.5 while iPTF14ajq peaks at -16.4, lending further support that these are high-energy events originating from very massive progenitors. It is noteworthy that the outliers we identified are similar not just in their early-time behavior but also in their later evolution, despite having been selected based on the first 30 days of their light curves. The outliers may also be of the same type identified in \cite{gonzalez-gaitan_rise-time_2015} as the long-rise population. They attribute these events to interloping IIn or 1987-like events, but this is not the case for our events. Clearly this population requires further study. 

\acknowledgements
We thank Y. Cao for reducing the spectrum of iPTF13blq. We thank A. Tanay and B. Zackay for helpful discussions. A.G.Y. is supported by the EU/FP7 via ERC grant no. 307260, the Quantum Universe I-CORE Program by the Israeli Committee for Planning and Budgeting and the Israel Science Foundation (ISF); by Minerva and ISF grants; by the Weizmann-UK ``making connections'' program; and by
Kimmel and ARCHES awards.

\bibliographystyle{apj}
\bibliography{biblio.bib}


\begin{appendix}

\begin{figure}[ht]
\centering
\includegraphics[width=0.45\columnwidth]{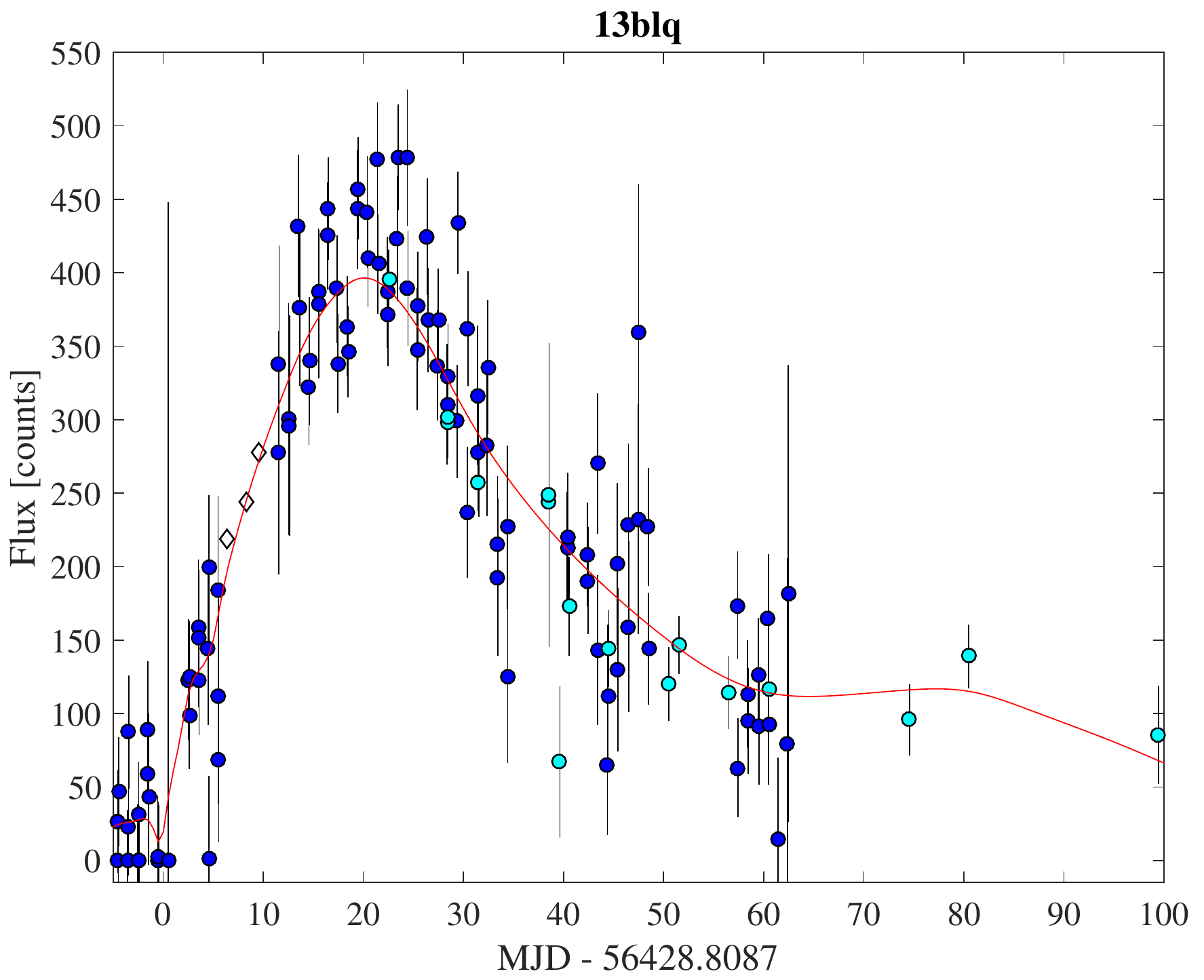}
\includegraphics[width=0.45\columnwidth]{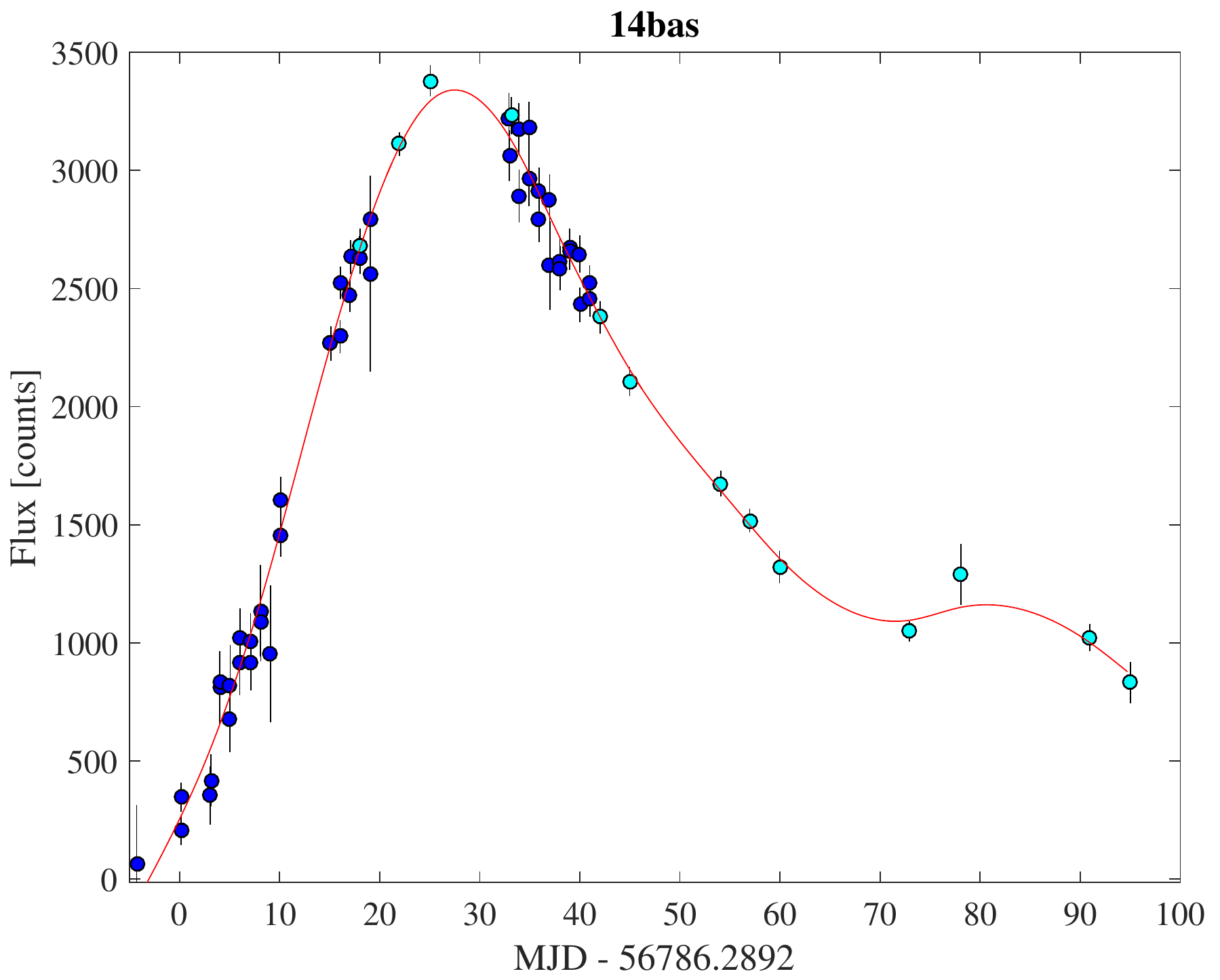}
\caption{Smoothed LCs of iPTF13blq and iPTF14bas. iPTF13blq was smoothed with the algorithm of \cite{rubin_type_2015} while iPTF14bas was smoothed with a smoothing spline.}
\label{fig:13blq_14bas_smoothed}
\end{figure}

\begin{figure}[ht]
\centering
\includegraphics[width=0.45\textwidth]{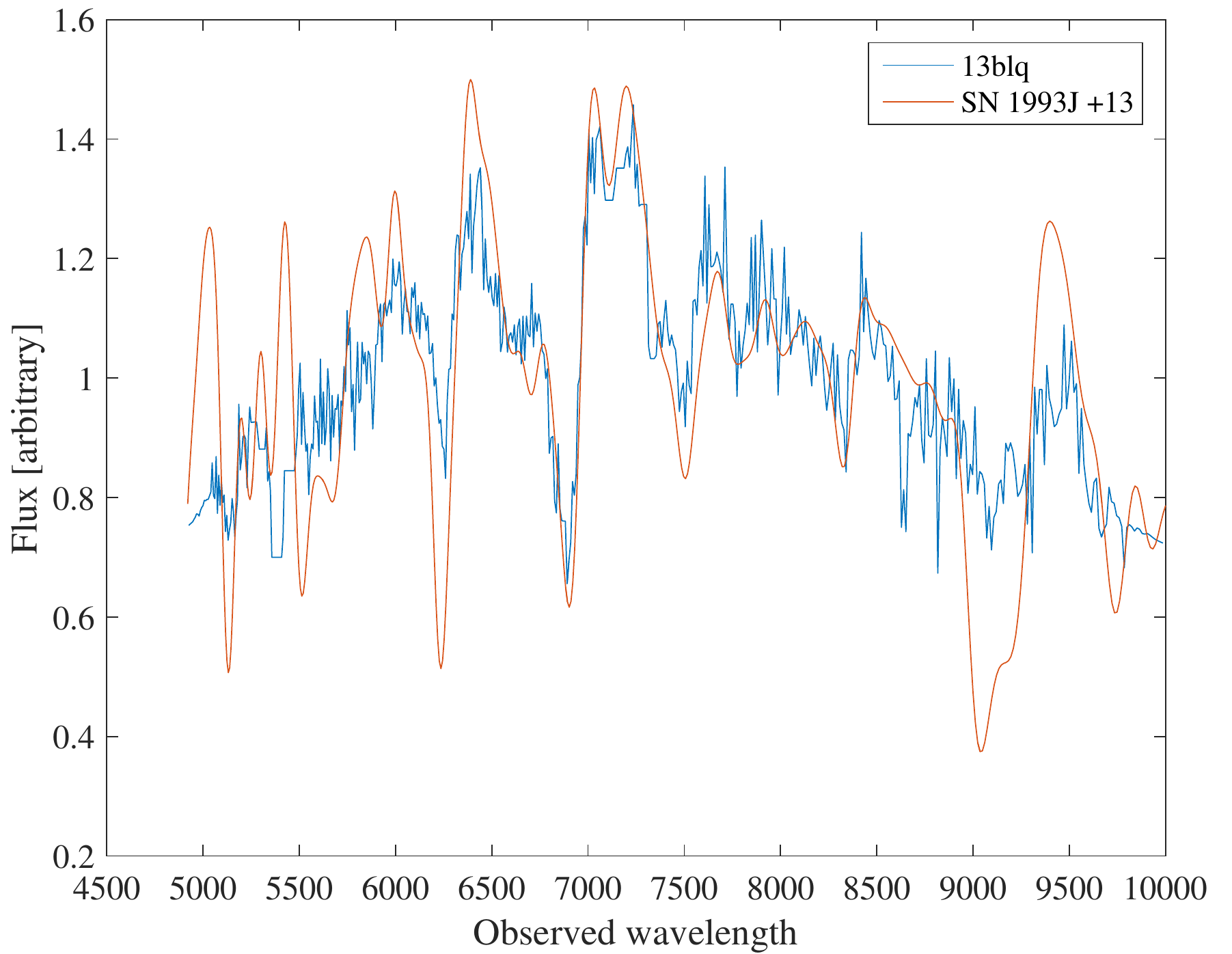}
\includegraphics[width=0.45\textwidth]{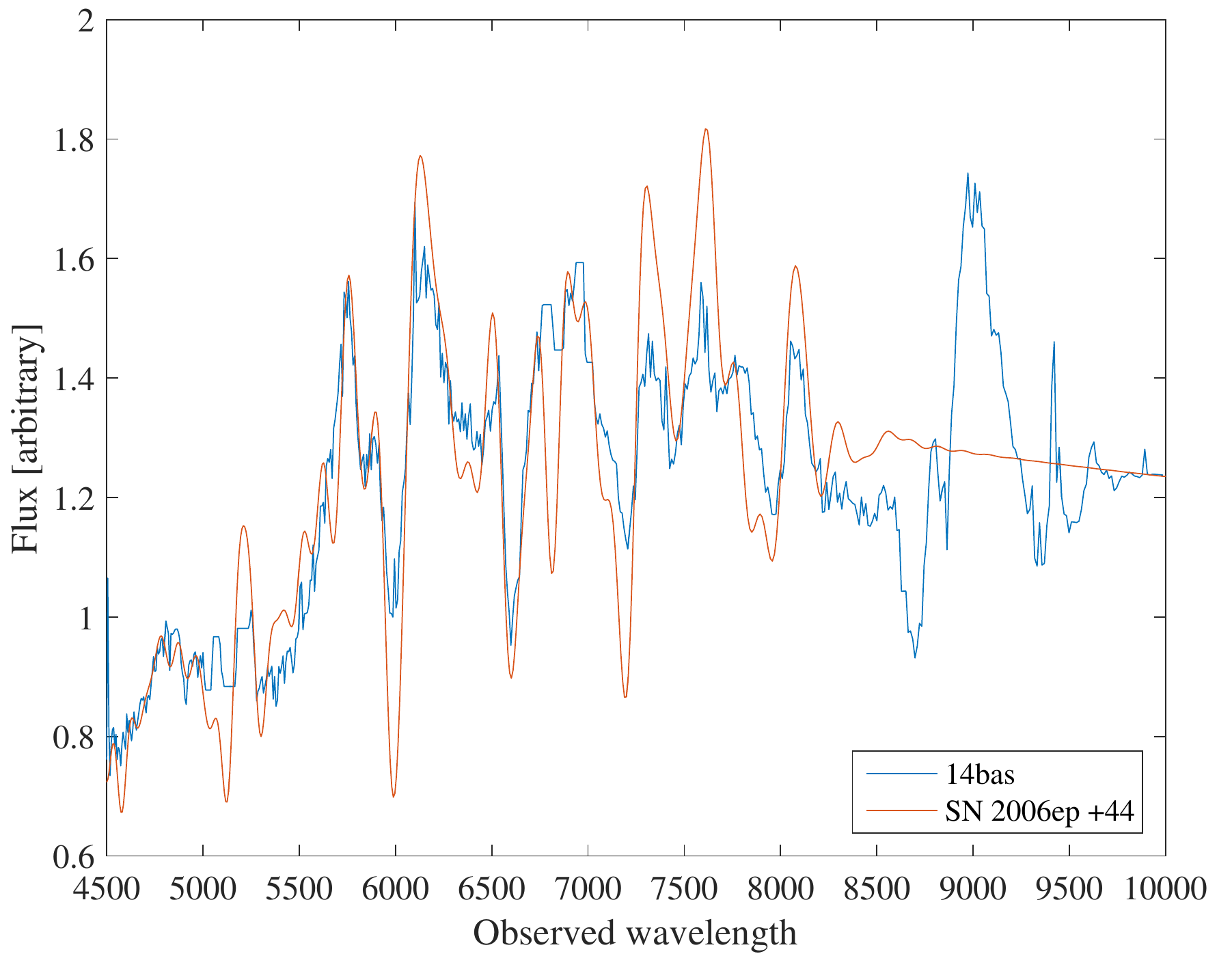}
\caption{SNID comparison of SNe iPTF13blq (left) and iPTF14bas (right) with IIb spectra.}
\label{fig:II-E_Classification}
\end{figure}
\end{appendix}

\end{document}

%% file: meanStdTablePreview.txt
1.0 & 0.68 & 0.22 & 0.68 & 0.28 & 1.54 & 0.40 \\ 
$\cdots$ & $\cdots$ & $\cdots$ & $\cdots$ & $\cdots$ & $\cdots$ & $\cdots$ \\ 
30.0 & -0.01 & 0.07 & 0.31 & 0.11 & 0.20 & 0.18 